\documentclass[3p,times,11pt,fleqn,sort&compress]{elsarticle}
\usepackage{graphicx,xcolor,amsmath,amssymb,amsthm,eufrak,physics,fullpage,mathrsfs}
\usepackage[hidelinks,colorlinks=true,linkcolor=blue,citecolor=blue,urlcolor=blue]{hyperref}
\usepackage[english]{babel}

\newcommand{\eq}[2]{\begin{gather}#2  \label{#1} \end{gather}}
\newcommand{\nn}{\nonumber\\}\newcommand{\Ra}{\mathrm{Ra}}\newcommand{\Bi}{\mathrm{Bi}}
\newcommand{\Rm}{\mathrm{Rm}}
\renewcommand{\order}[1]{\mathcal{O}\big(#1~\!\big)}

\journal{Journal}

\begin{document}

\begin{frontmatter}

\title{\bf The Wooding problem revisited} 

\author[1]{A. Barletta\corref{cor1}} 
\author[2]{D. A. S. Rees}

\address[1]{Department of Industrial Engineering, Alma Mater Studiorum Università di Bologna,\\ Viale Risorgimento 2, Bologna 40136, Italy}
\address[2]{Department of Mechanical Engineering, University of Bath, Claverton Down, BA2 7AY Bath, United Kingdom}
\cortext[cor1]{Corresponding author,~ \texttt{antonio.barletta@unibo.it}}


\begin{abstract}
The threshold conditions to convective instability in a semi-infinite porous layer saturated by a fluid are determined. The classical setup for this problem in geothermal fluid dynamics was originally modelled by Wooding in 1960. Its formulation is here reconsidered to allow for an imperfect heat transfer across the boundary, parametrised through the Biot number. The temperature boundary condition considered by Wooding is here recovered as the limit of an infinite Biot number. The linear stability analysis of the stationary boundary layer which establishes in the porous medium when a boundary steady suction occurs is carried out. Two different versions of the Rayleigh number are considered, namely, a temperature-difference-based version and a heat-flux-based version. While the former is the classical Rayleigh number for flow in porous media, the latter is a variant definition which displays a finite limit at neutral stability in both the opposite limiting cases of an infinite or of a zero Biot number.
\end{abstract}
\begin{keyword}
Thermal instability \sep Convection heat transfer \sep Porous medium \sep Boundary layer \sep Darcy's law \sep Buoyancy-induced flow



\end{keyword}

\end{frontmatter}

\section{Introduction}
The analysis carried out in the pioneering paper by \citet{wooding1960rayleigh} provides an interesting model of the thermal convection currents of the groundwater saturating a semi-infinite porous medium below a lake in the geothermal region of Wairakei in New Zealand. The porous medium is subject to a vertical temperature gradient, while the horizontal interface between the porous medium and the overlying water is exposed to a suction velocity condition. As pointed out by the author, ``when a steady state has been established, a thermal boundary
layer of exponential form then exists below the surface''. The thermal boundary layer can develop a linear instability when the groundwater thermal gradient is high enough or, parametrically, when the Rayleigh number (or Darcy-Rayleigh number) exceeds a critical threshold $\Ra_c = 14.35$ with a dimensionless wave number of the normal mode equal to $k_c = 0.759$ \cite{van2001stability}.

More recent studies of the steady thermal boundary layer in a semi-infinite porous medium were carried out by \citet{rees2009onset}, \citet{patil2013linear}, \citet{rees2018inclined} and by \citet{islam2024onset}. These authors extended the analysis of the Wooding problem by accounting for the nonlinear dynamics of large amplitude perturbations \cite{rees2009onset}, for the lack of local thermal equilibrium between solid and fluid phases \cite{patil2013linear}, for the inclination of the plane boundary of the porous medium relative to the horizontal \cite{rees2018inclined} and for the anisotropy of the porous medium \cite{islam2024onset}.

There has been a very wide interest in the last sixty years about the onset of the Rayleigh-B\'enard instability in a porous layer heated from below and subject to injection-suction boundary conditions for the seepage velocity \cite{sutton1970onset, homsy1976convective, jones1986convective}. This instability problem proposed by \citet{sutton1970onset} displays an extra parameter with respect to the Wooding problem. That is due to the finite thickness of the porous layer, $H$, leading to the definition of a P\'eclet number given by the ratio between the injection-suction velocity at the boundaries, $V_0$, and the diffusivity-induced velocity, $\alpha/H$, where $\alpha$ is the average thermal diffusivity of the saturated porous medium. In fact, the critical values $(k_c, \Ra_c)$ for the Wooding problem were recovered by \citet{homsy1976convective} in the limit of an infinite P\'eclet number. As pointed out by \citet{homsy1976convective}, for large values of the thickness $H$, a boundary layer structure develops so that the porous medium ``becomes approximately isothermal except for a thermal layer'' whose thickness is of the order of magnitude of $\alpha/V_0$. Thus, the constant $\alpha/V_0$ turns out to be the natural reference length for the dimensional analysis of the Wooding problem \cite{homsy1976convective}. The nonlinear effects and the subcriticality of the transition to instability in Sutton's problem \cite{sutton1970onset} have been thoroughly investigated in the recent papers by \citet{capone2023weakly} and by \citet{gianfrani2025eckhaus}. Furthermore, \citet{deepika2017onset} considered the effects of Brinkman's term in the momentum balance for the porous medium, \citet{turkyilmazoglu2023darcy} investigated different combinations of Dirichlet and Neumann conditions at the boundaries, while \citet{capone2024throughflow} studied the linear and nonlinear instability for an extended Sutton's problem based on the model of bidisperse flow in a saturated porous medium.

The aim of this paper is an extended study of the Wooding problem where heat transfer to the fluid reservoir overlying the semi-infinite porous medium is modelled by a finite thermal resistance or, in nondimensional terms, by a finite Biot number. The study of an imperfect heat transfer at the boundary of a semi-infinite porous medium was previously carried out by \citet{hitchen2016impact} although in a different framework. The boundary was assumed to be impermeable and subject to a Robin boundary condition for the temperature \cite{hitchen2016impact}. Such assumptions lead the authors to a basic thermal boundary layer solution explicitly time-dependent, unlike the stationary basic state considered by Wooding \cite{wooding1960rayleigh}, as well as in our analysis. In fact, the time-independence in the basic thermal boundary layer considered in the Wooding problem is a consequence of the vertical throughflow velocity prescribed at the upper boundary of the semi-infinite porous medium. Our investigation of the extended Wooding problem involves, in addition to the Biot number, a two-fold definition of a Rayleigh number and a modified Rayleigh number, where the latter has the advantage to include regular asymptotic solutions for the linear stability analysis in both the limiting cases of a zero or infinite Biot number. The neutral stability condition for the extended Wooding problem, together with the values of the critical wavenumber and Rayleigh number, is obtained numerically by spanning different values of the Biot number.

\section{The problem formulation}
\begin{figure}[t]
\centering
\includegraphics[width=0.5\textwidth]{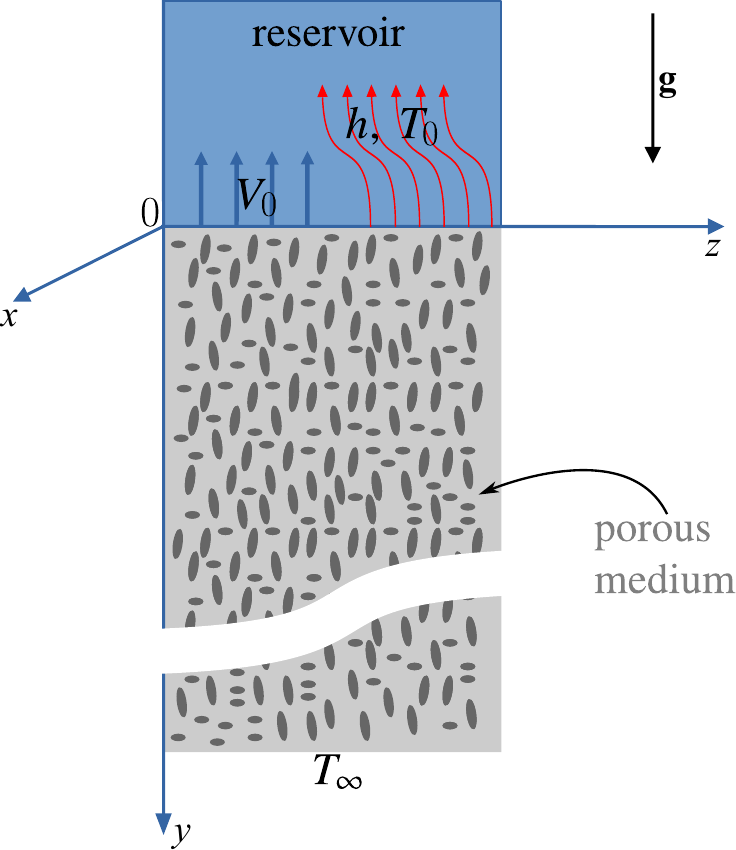}
\caption{\label{fig1}A sketch of the semi-infinite porous medium and its boundary conditions.}
\end{figure}

Let us consider a semi-infinite porous medium saturated by a Newtonian fluid. The $y$ axis is chosen as vertical and downward oriented, while the $x$ and $z$ axes are horizontal (see Fig.~\ref{fig1}). Thus, the gravitational acceleration is expressed as $\vb{g} = g\, \vu{e}_y$, where $g$ is its modulus and $\vu{e}_y$ is the $y$ axis unit vector. The porous medium is considered as isotropic and homogeneous, with local thermal equilibrium between the solid and the fluid phases. The medium permeability, $K$, is considered as sufficiently small so that Darcy's law can be applied to express the local momentum balance \citep{NieldBejan2017}.

\subsection{The boundary and asymptotic conditions}
As sketched in Fig.~\ref{fig1}, the semi-infinite porous medium is bounded by the horizontal plane $y=0$, while it extends unboundedly over the region $y > 0$. The saturated porous medium is cooled from above by thermal contact with a fluid reservoir having an undisturbed temperature $T_{0}$ which is smaller than the temperature $T_{\infty}$ reached asymptotically by the medium in the limit $y \to +\infty$. The heat transfer between the boundary $y=0$ and  the upper reservoir is modelled via a Robin temperature condition with a constant heat transfer coefficient, $h$. As the reservoir has a pressure smaller than that of the semi-infinite porous medium, there is a uniform and constant suction velocity $V_0>0$ prescribed at $y=0$. Thus, the boundary conditions can be expressed as
\eq{1}{
y = 0 : \qquad v = - V_0 \qc \chi \pdv{T}{y} - h \qty( T - T_0) = 0 ,
}
where we denoted with $\qty(u,v,w)$ the Cartesian components of the seepage velocity, with $\chi$ the average thermal conductivity of the saturated porous medium and with $T$ the temperature field. On the other hand, the conditions at infinity are given by  
\eq{2}{
y \to +\infty : \qquad u = 0 = w \qc T = T_{\infty} ,
}
where $u=0=w$ defines an asymptotic condition where the fluid undergoes no horizontal flow, or, no horizontal pressure gradient exists.

\subsection{The governing equations}
The local mass, momentum and energy balance equations can be expressed by claiming the validity of the Oberbeck-Boussinesq approximation under the assumption that viscous dissipation effects are negligible,
\eq{3}{
\pdv{u}{x} + \pdv{v}{y} + \pdv{w}{z} = 0, \nn
\frac{\mu}{K} u = - \pdv{p}{x} \qc \frac{\mu}{K} v = - \pdv{p}{y} + \rho_{\infty} g \beta (T_\infty - T) \qc \frac{\mu}{K} w = - \pdv{p}{z}, \nn
\xi \pdv{T}{t} + u \pdv{T}{x} + v \pdv{T}{y} + w \pdv{T}{z} = \alpha \laplacian{T} ,
}
where $t$ is time, $p$ is the difference between the pressure and the hydrostatic pressure, $\mu$ is the fluid viscosity, $\alpha$ is the average thermal diffusivity of the saturated medium and $\xi$ is the ratio between the average heat capacity of the saturated medium and that of the fluid. Furthermore, $\rho_{\infty}$ is the reference fluid density evaluated at $T=T_{\infty}$ and $\beta$ is the thermal expansion coefficient of the fluid.

\subsection{Dimensionless formulation}
Following the scaling proposed by \citet{rees2018inclined}, one can define the reference length, the reference time and the reference temperature difference
\eq{4}{
\ell_0 = \frac{\alpha}{V_0} \qc t_0 = \frac{\xi\alpha}{V_0^2} \qc \Delta T = T_\infty - T_0 .
}
Hence, one can determine the dimensionless quantities, identified by asterisks,
\eq{5}{
\frac{(x,y,z)}{\ell_0} = (x^*,y^*,z^*) \qc \frac{(u,v,w)}{V_0} = (u^*,v^*,w^*) \qc \frac{t}{t_0} = t^* , \nn 
\frac{p}{\mu\alpha/K}=p^* \qc \frac{T_{\infty} - T}{\Delta T} = T^*,
}
and the dimensionless parameters
\eq{6}{
\Ra = \frac{\rho_{\infty} g \beta \Delta T K \ell_0}{\alpha \mu} \qc \Bi = \frac{h \ell_0}{\chi} ,
}
{\em i.e.}, the Rayleigh number for a saturated porous medium and the Biot number, respectively. By utilising \eqref{4}-\eqref{6}, one can express the dimensionless version of the governing equations \eqref{3} and of the boundary and asymptotic conditions \eqref{1} and \eqref{2} as
\eq{7}{
\pdv{u}{x} + \pdv{v}{y} + \pdv{w}{z} = 0, \nn
u = - \pdv{p}{x} \qc v = - \pdv{p}{y} + \Ra T \qc w = - \pdv{p}{z}, \nn
\pdv{T}{t} + u \pdv{T}{x} + v \pdv{T}{y} + w \pdv{T}{z} = \laplacian{T} , \nn
y = 0 : \hspace{12mm} v = - 1 \qc \pdv{T}{y} - \Bi T = - \Bi ,\nn
y \to +\infty : \qquad u = 0 = w \qc T = 0 ,
}
where we have suppressed the asterisks for the sake of brevity. 

\subsection{Stationary base solution}
A stationary solution of \eqref{7} exists where the fields $\qty(u,v,w)$, $T$ and $p$ are given by
\eq{8}{
\qty(u_b, v_b, w_b) = \qty(0, -1, 0) \qc T_b = \frac{\Bi}{\Bi + 1}\, e^{-y} \qc p_b = y - \frac{\Ra\, \Bi}{\Bi + 1}\,  e^{-y} + constant.
}
In \eqref{8}, the subscript $b$ indicates the basic solution. In the limit of an infinite Biot number, \eqref{8} agrees with the basic solution considered by \citet{islam2024onset}.

\section{Linear stability analysis}
The model defined by \eqref{7} and the base solution defined by \eqref{8} are invariant under rotations around the $y$ axis. In other words, all the wavelike perturbations of the base solution that propagate along any horizontal direction are equivalent with respect to the linear stability analysis. Then, without any loss of generality, one can formulate the study as relative to two-dimensional $z$-independent perturbation modes which propagate along the $x$ axis,
\eq{9}{
\qty(u,v,w) = \qty(0, -1, 0) + \epsilon\, \qty(\pdv{\psi}{y}, - \pdv{\psi}{x} , 0) \qc T = \frac{\Bi}{\Bi + 1}\, e^{-y} + \epsilon\, \theta .
}
Here, $\psi\qty(x,y,t)$ and $\theta\qty(x,y,t)$ are the perturbation streamfunction and temperature, respectively, while $\epsilon$ is a small perturbation parameter, $|\epsilon| \ll 1$.

We point out that, in order to fulfil the condition that $\epsilon\, |\theta|$ be smaller than $T_b$ for every $y>0$, the condition,
\eq{9b}{
\lim_{y \to +\infty} e^y\, |\theta(x,y,t)| < \infty,
}
must be satisfied. Such a condition not only implies that $\theta(x,y,t) \to 0$ when $y \to +\infty$, but also that it decays with $y$, at least, as fast as $e^{-y}$.

\subsection{Streamfunction-temperature formulation}

Substitution of \eqref{9} into \eqref{7} leads to a linearised streamfunction-temperature formulation obtained by evaluating the curl of vector $\qty(u,v,w)$ and by neglecting terms of order $\epsilon^2$,
\eq{10}{
\laplacian_{(x,y)} \psi + \Ra \pdv{\theta}{x} = 0, \nn
\laplacian_{(x,y)} \theta - \pdv{\theta}{t} + \pdv{\theta}{y} - \frac{\Bi}{\Bi + 1}\, e^{-y} \pdv{\psi}{x} = 0 ,\nn
y = 0 : \hspace{12mm} \pdv{\psi}{x} = 0 \qc \pdv{\theta}{y} - \Bi \theta = 0 ,\nn
y \to +\infty : \qquad \pdv{\psi}{y} = 0  \qc \theta = 0 ,
}
where $\laplacian_{(x,y)}$ is the two-dimensional Laplacian operator in the $(x,y)$ plane. The first equation \eqref{7} is identically satisfied by the streamfunction definition and, hence, it gives no contribution to \eqref{10}.

\subsection{Normal modes}

The homogeneous differential problem \eqref{10} can be studied by employing normal modes, given by
\eq{11}{
\psi\qty(x,y,t) = \Psi(y)\, e^{\lambda t} \cos(k x) \qc \theta\qty(x,y,t) = \Theta(y)\, e^{\lambda t} \sin(k x),
}
where $k$ is the wavenumber and $\lambda = \sigma - i \omega$ is a complex variable such that its real part, $\sigma$, is the perturbation time-growth rate and $\omega$ is the mode angular frequency. When $\sigma > 0$ we have instability, when $\sigma < 0$ we have stability and when $\sigma = 0$ we have neutral stability.
Then, \eqref{10} can be turned into an ordinary differential problem,
\eq{12}{
\Psi'' - k^2 \Psi + k \Ra \Theta = 0, \nn
\Theta'' - \qty(k^2 + \lambda) \Theta + \Theta' + \frac{k\, \Bi}{\Bi + 1}\, e^{-y}\ \Psi = 0 ,\nn
y = 0 : \hspace{12mm} \Psi = 0 \qc \Theta' - \Bi\, \Theta = 0 ,\nn
y \to +\infty : \qquad \Psi' = 0  \qc \Theta = 0 ,
}
where primes denote derivatives with respect to $y$. 

\subsection{Exchange of stabilities}

We follow the same procedure presented in \citet{islam2024onset} in order to prove the validity of the principle of exchange of stabilities. In particular, we define function 
\eq{13}{
\Phi(y) = e^{y/2}\ \Theta(y),
}
so that \eqref{12} is given by
\eq{14}{
\Psi'' - k^2 \Psi + k \Ra\, e^{-y/2}\ \Phi = 0, \nn
\Phi'' - \qty(k^2 + \lambda + \frac{1}{4}) \Phi + \frac{k\, \Bi}{\Bi + 1}\, e^{-y/2}\ \Psi = 0 ,\nn
y = 0 : \hspace{12mm} \Psi = 0 \qc \Phi' - \qty(\Bi + \frac{1}{2}) \Phi = 0 ,\nn
y \to +\infty : \qquad \Psi' = 0  \qc \Phi = 0 .
}
We point out that the condition $\Phi(y) \to 0$ when $y \to +\infty$ is a consequence of \eqref{9b} and of the definitions \eqref{11} and \eqref{13}.
Thus, one can write
\eq{15}{
\Psi\bar\Psi'' - k^2 \Psi\bar\Psi + k \Ra\, e^{-y/2}\ \Psi\bar\Phi = 0, \nn
\bar\Phi\Phi'' - \qty(k^2 + \lambda + \frac{1}{4}) \bar\Phi\Phi + \frac{k\, \Bi}{\Bi + 1}\, e^{-y/2}\ \bar\Phi\Psi = 0 ,
}
where the overline denotes complex conjugation.
If we integrate both equations \eqref{15} with respect to $y$, we obtain
\eq{16}{
\int_0^{+\infty} \Psi\bar\Psi''\ \dd y - k^2 \int_0^{+\infty} \qty|\Psi|^2\ \dd y = - k \Ra \int_0^{+\infty}  e^{-y/2}\ \Psi\bar\Phi\ \dd y , \nn
\int_0^{+\infty} \bar\Phi\Phi''\ \dd y - \qty(k^2 + \sigma + \frac{1}{4}) \int_0^{+\infty} \qty|\Phi|^2\ \dd y + \frac{k\, \Bi}{\Bi + 1} \int_0^{+\infty}  e^{-y/2}\ \bar\Phi\Psi\ \dd y  \nn
\hspace{72mm}= - i \omega \int_0^{+\infty} \qty|\Phi|^2\ \dd y  .
}
Integrations by parts yield
\eq{17}{
\int_0^{+\infty} \qty|\Psi'|^2\ \dd y + k^2 \int_0^{+\infty} \qty|\Psi|^2\ \dd y = k \Ra \int_0^{+\infty}  e^{-y/2}\ \Psi\bar\Phi\ \dd y , 
}
and
\eq{18}{
\int_0^{+\infty} \qty|\Phi'|^2\ \dd y + \qty(\Bi + \frac{1}{2}) \qty|\Phi(0)|^2 + \qty(k^2 + \sigma + \frac{1}{4}) \int_0^{+\infty} \qty|\Phi|^2\ \dd y \nn
\hspace{5cm}- \frac{k\, \Bi}{\Bi + 1} \int_0^{+\infty}  e^{-y/2}\ \Psi\bar\Phi\ \dd y  = i \omega \int_0^{+\infty} \qty|\Phi|^2\ \dd y  .
}
Use has been made of the conditions at $y = 0$ and $y \to +\infty$ given by \eqref{14}. In particular, such conditions lead to
\eq{19}{
\qty[\bar\Phi\Phi']_0^{+\infty} = \lim_{y \to +\infty} \bar\Phi(y)\Phi'(y) - \qty(\Bi + \frac{1}{2}) \qty|\Phi(0)|^2 = - \qty(\Bi + \frac{1}{2}) \qty|\Phi(0)|^2 .
}
Equation \eqref{17} shows that the integral 
\eq{20}{
\int_0^{+\infty}  e^{-y/2}\ \Psi\bar\Phi\ \dd y
}
is real, so that one infers that the left hand side of \eqref{18} is also real. Then, since the right hand side of \eqref{18} is purely imaginary or zero, the conclusion is that either $\omega = 0$ or $\Phi(y)$ is identically zero. The latter option is to be excluded because it would imply, from \eqref{14}, that also $\Psi(y)$ is identically zero or, equivalently, that no disturbance perturbs the basic state. Thus, $\omega$ must be zero and the principle of exchange of stabilities holds. 

\subsection{The limiting case of small Biot numbers}\label{smallbi}
On exploring the regime where the Biot number is very small and eventually drops to zero, one may introduce a useful reformulation of the governing equations and boundary conditions \eqref{12} based on the definitions
\eq{21}{
\hat\Theta = \frac{\Bi + 1}{\Bi}\, \Theta \qc \Rm = \frac{\Bi}{\Bi + 1} \,\Ra,
}
so that \eqref{12} can be rewritten as,
\eq{22}{
\Psi'' - k^2 \Psi + k\, \Rm\, \hat\Theta = 0, \nn
\hat\Theta'' - \qty(k^2 + \lambda) \hat\Theta + \hat\Theta' + k\, e^{-y}\ \Psi = 0 ,\nn
y = 0 : \hspace{12mm} \Psi = 0 \qc \hat\Theta' - \Bi\, \hat\Theta = 0 ,\nn
y \to +\infty : \qquad \Psi' = 0  \qc \hat\Theta = 0 ,
}
where $\Rm$ is a modified Rayleigh number. Physically speaking, the construction of $\Rm$ given by \eqref{21} captures a situation where the dimensional temperature difference $\Delta T$ defined by \eqref{4} is replaced with 
\eq{23}{
\Delta T_{\rm m} = \frac{q_0 \ell_0}{\chi} \qc \qfor q_0 = \frac{h \qty(T_{\infty} - T_0)}{\Bi + 1} ,
}
in the definition of Rayleigh number. In \eqref{23}, the quantity $q_0$ is a dimensional heat flux, namely a thermal power per unit area. In other words, using $\Rm$ instead of $\Ra$ serves to capture an isoflux thermal condition occurring at the boundary $y=0$ in the limit $\Bi \to 0$,
so that \eqref{22} yields
\eq{24}{
\Psi'' - k^2 \Psi + k\, \Rm\, \hat\Theta = 0, \nn
\hat\Theta'' - \qty(k^2 + \lambda) \hat\Theta + \hat\Theta' + k\, e^{-y}\ \Psi = 0 ,\nn
y = 0 : \hspace{12mm} \Psi = 0 \qc \hat\Theta' = 0 ,\nn
y \to +\infty : \qquad \Psi' = 0  \qc \hat\Theta = 0 .
}

\section{Discussion of the results}

\begin{figure}[t]
\centering
\includegraphics[height=65mm]{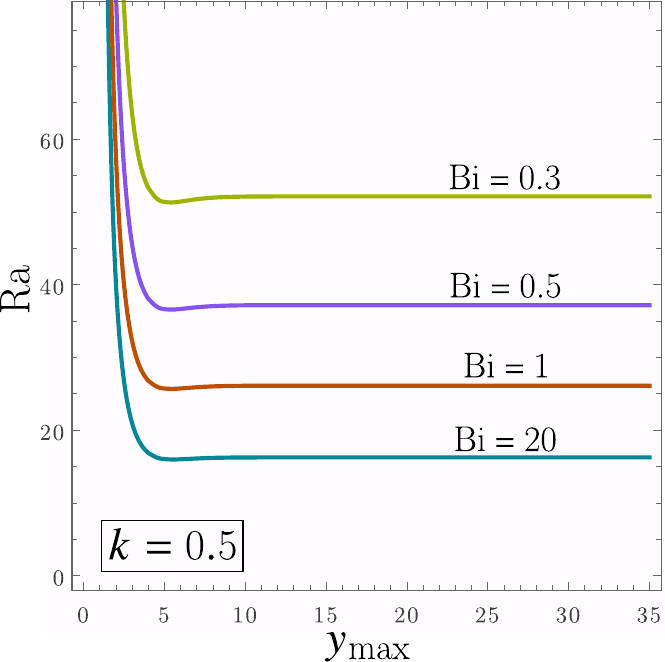}(a)\quad\includegraphics[height=65mm]{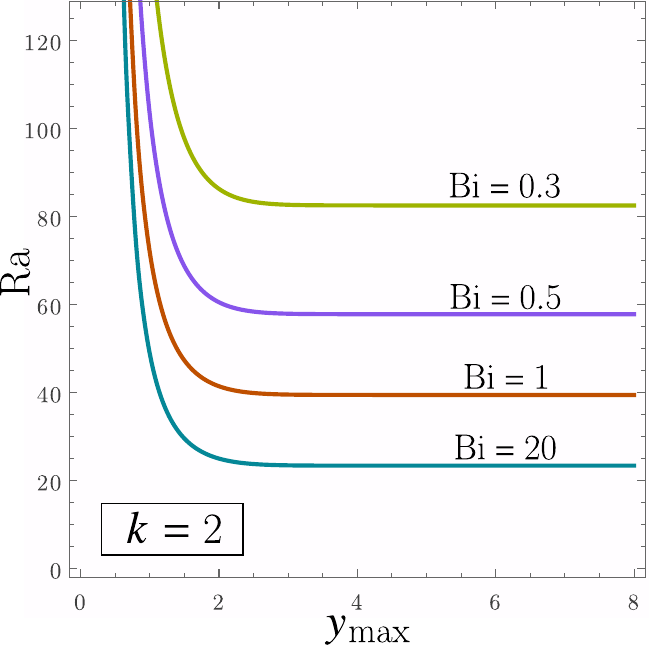}(b)
\caption{\label{fig2}Neutral stability values of $\Ra$ versus $y_{\max}$ for $k=0.5$ (a) and for $k=2$ (b), with different Biot numbers.}
\end{figure}

\subsection{Numerical method}
The solution of the eigenvalue problem given by \eqref{12} or \eqref{22} can be obtained numerically by employing the shooting method. A detailed description of the numerical method and its application for the solution of the stability eigenvalue problems is provided in \citet{Straughan} and in \citet{barletta2019routes}. The implementation of this method into an {\sl Octave} \citep{octave} code is described in details in Chapter 10 of \citet{barletta2019routes}. A description of the shooting method applied to a similar problem, {\em i.e.} the Wooding problem for an inclined boundary, is also available in \citet{rees2018inclined}. As pointed out by these authors \citep{rees2018inclined}, a special feature of the Wooding problem is that it is relative to a semi-infinite domain. Then, unlike most analyses of thermal instability, the target conditions for the shooting method are not prescribed at a predefined value of $y$, but they are given as limiting conditions when $y \to +\infty$. In order to set up reliable numerical solutions, one has to artificially locate the target conditions at $y = y_{\max}$ and then to test what happens on gradually increasing the value of $y_{\max}$. 

Figure~\ref{fig2} illustrates the effect of an increasing value of $y_{\max}$ on the neutral stability $(\lambda = 0)$ value of $\Ra$ for two wavenumbers, $k=0.5,2$, and for different values of $\Bi$. It is evident from Fig.~\ref{fig2} that a reliable numerical implementation of the  asymptotic condition for $y \to +\infty$ can be obtained with a finite $y_{\max}$ whose value markedly depends on the prescribed wavenumber. In particular, Fig.~\ref{fig2} suggests that larger $y_{\max}$ are needed for $k=0.5$ than 
for $k=2$. This check is very important for the numerical determination of the neutral stability curves in the $(k, \Ra)$ plane and in the $(k, \Rm)$ plane. The numerical results presented in the forthcoming discussion are obtained by testing their sensitivity to the assigned value of $y_{\max}$. In practice, a sufficiently larger value is used for $y_{\max}$, such as 15 or 20.

\begin{figure}[t]
\centering
\includegraphics[height=65mm]{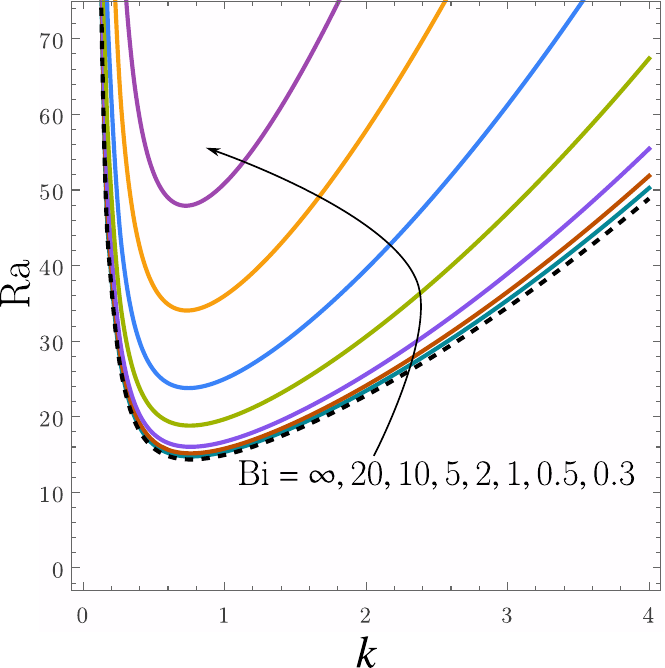}(a)\quad\includegraphics[height=65mm]{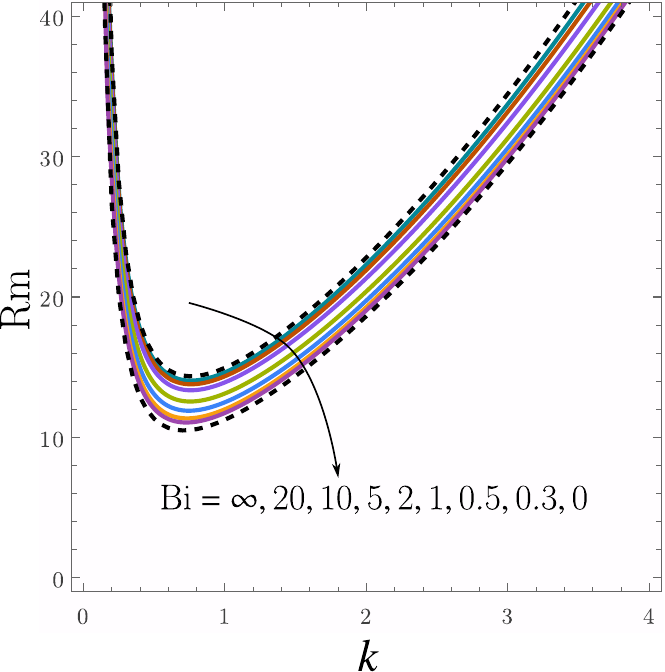}(b)
\caption{\label{fig3}Neutral stability curves in the $\qty(k,\Ra)$ plane (a) and in the $\qty(k,\Rm)$ plane (b), with different Biot numbers.}
\end{figure}

\subsection{Neutral stability curves}
The neutral stability curves $(\lambda = 0)$ for different Biot numbers can be either represented in the $(k, \Ra)$ plane or in the $(k, \Rm)$ plane. Their appearence is deeply different in the two cases, as one may see in Fig.~\ref{fig3}. This is a consequence of the choice of the transition parameter for the onset of instability. When basing the analysis on the Rayleigh number, $\Ra$, then the asymptotic case where $\Bi \to 0$ becomes one of stability. More precisely, one can say that, with $\Bi \to 0$, the basic solution \eqref{8} is linearly stable for every given $\Ra$, whatever is the wavenumber $k$ of the normal mode perturbation. Physically, this means that one can increase the dimensional temperature difference $\Delta T$, defined by \eqref{4}, without ever observing any instability. This makes perfect sense, because $\Bi \to 0$ is an asymptotic case where no thermal interaction between the underlying porous medium and the overlying fluid is possible. This feature can be easily retrieved also from \eqref{8}, which evidently shows that $\Bi \to 0$ yields a case where the porous medium temperature remains perfectly uniform.

On the other hand, with $\Bi \to 0$, linear instability is possible provided that the value of $\Rm$ is sufficiently large. The reason of this dissimilarity is that $\Rm$ is a Rayleigh number based on the dimensional temperature difference $\Delta T_{\rm m}$, defined by \eqref{23}. As already mentioned in Section~\ref{smallbi}, $\Delta T_{\rm m}$ is a heat flux based quantity where the dimensional heat flux $q_0$ remains finite and nonzero even when $\Bi \to 0$. Indeed, if one applies the scaling factor $\qty(\Bi + 1)/\Bi$ not only to the temperature perturbation, as shown in \eqref{21}, but also to the basic temperature $T_b$, given by \eqref{8}, then the basic temperature gradient exponentially decaying in $y$ remains active also when $\Bi \to 0$.

The utterly different parametrisations for the transition to instability, one based on $\Ra$ and the other one on $\Rm$, underpin different physical descriptions. Figure~\ref{fig3} shows that, according to the two descriptions, the Biot number acts as a destabilising parameter, if one relies on the $\Ra$-parametrisation, or as a stabilising parameter, if one relies on the $\Rm$-parametrisation. 

\subsection{Critical values}
In order to obtain the critical values of $\Ra$ or $\Rm$ for a prescribed Biot number, we have to minimise such parameters versus $k$ along the neutral stability curve. The minimisation procedure can be carried out efficiently by doubling the differential order of the eigenvalue problem. With no loss of generality, we rely on the $\Rm$-parametrisation and consider \eqref{22}, with $\lambda=0$. The doubled order system is
\eq{25}{
\Psi'' - k^2 \Psi + k\, \Rm\, \hat\Theta = 0, \nn
\hat\Theta'' - k^2 \hat\Theta + \hat\Theta' + k\, e^{-y}\ \Psi = 0 ,\nn
\Psi_k'' - k^2 \Psi_k + k\, \Rm\, \hat\Theta_k - 2 k\, \Psi + \Rm\, \hat\Theta = 0, \nn
\hat\Theta_k'' - k^2 \hat\Theta_k + \hat\Theta_k' + k\, e^{-y}\ \Psi_k - 2 k\, \hat\Theta + e^{-y}\ \Psi = 0 ,\nn
y = 0 : \hspace{12mm} \Psi = 0 \qc \hat\Theta' - \Bi \hat\Theta = 0 \qc \Psi_k = 0 \qc \hat\Theta_k' - \Bi\, \hat\Theta_k = 0 ,\nn
y \to +\infty : \qquad \Psi' = 0  \qc \hat\Theta = 0 \qc \Psi_k' = 0  \qc \hat\Theta_k = 0 ,
}
where $\Psi_k = \partial \Psi/\partial k$, $\hat\Theta_k = \partial \hat\Theta/\partial k$. The third and fourth differential equations have been obtained by deriving the first and second differential equations with respect to $k$. We have implicitly assumed that $\partial \Rm/\partial k = 0$ and $\partial \Bi/\partial k = 0$, where the former is a consequence of $\Rm(k)$ being at its minimum in order to detect the critical value, while the second is a consequence of $\Bi$ being a prescribed input parameter in the solution. The numerical solution of the doubled eigenvalue problem \eqref{25} for a given Biot number leads to the determination of $\qty(k_c, \Rm_c)$ as eigenvalues and $\qty(\Psi, \hat\Theta, \Psi_k, \hat\Theta_k)$ as eigenfunctions.

\subsubsection{Critical values for small Biot numbers} 
An approximate solution of \eqref{25} can be obtained when $\Bi \ll 1$.  This task can be accomplished by writing the series expansions in powers of $\Bi$,
\eq{26}{
\Psi(y) = \Psi_0(y) + \Psi_1(y)\, \Bi + \order{\Bi^2} \qc \hat\Theta(y) = \hat\Theta_0(y) + \hat\Theta_1(y)\, \Bi + \order{\Bi^2} , \nn
\Psi_k(y) = \Psi_{k0}(y) + \Psi_{k1}(y)\, \Bi + \order{\Bi^2} \qc \hat\Theta_k(y) = \hat\Theta_{k0}(y) + \hat\Theta_{k1}(y)\, \Bi + \order{\Bi^2} , \nn
k = k_0 + k_1\, \Bi  + \order{\Bi^2} \qc \Rm = \Rm_0 + \Rm_1\, \Bi  + \order{\Bi^2} .
}
A scale fixing condition can be set as the eigenfunctions are defined only up to an arbitrary additive constant. We choose to prescribe
\eq{27}{
\hat\Theta(0) = 1 .
}
which implies $\hat\Theta_0(0) = 1$, $\hat\Theta_{k0}(0) = 0$, $\hat\Theta_1(0) = 0$ and $\hat\Theta_{k1}(0) = 0$, while the boundary conditions in \eqref{25} yield $\hat\Theta'_0(0) = 0$, $\hat\Theta'_{k0}(0) = 0$, $\hat\Theta'_1(0) = 1$ and $\hat\Theta'_{k1}(0) = 0$. Thus, substitution of \eqref{26} in \eqref{25} leads to a $0^{th}$-order differential problem
\eq{28}{
\Psi_0'' - k_0^2 \Psi_0 + k_0\, \Rm_0\, \hat\Theta_0 = 0, \nn
\hat\Theta_0'' - k_0^2 \hat\Theta_0 + \hat\Theta_0' + k_0\, e^{-y}\ \Psi_0 = 0 ,\nn
\Psi_{k0}'' - k_0^2 \Psi_{k0} + k_0\, \Rm_0\, \hat\Theta_{k0} - 2 k_0\, \Psi_0 + \Rm_0\, \hat\Theta_0 = 0, \nn
\hat\Theta_{k0}'' - k_0^2 \hat\Theta_{k0} + \hat\Theta_{k0}' + k_0\, e^{-y}\ \Psi_{k0} - 2 k_0\, \hat\Theta_0 + e^{-y}\ \Psi_0 = 0 ,\nn
y = 0 : \hspace{12mm} \Psi_0 = 0 \qc \hat\Theta_0 = 1 \qc \hat\Theta'_0 = 0 \qc \Psi_{k0} = 0 \qc \hat\Theta_{k0} = 0 \qc \hat\Theta'_{k0} = 0 ,\nn
y \to +\infty : \qquad \Psi_0' = 0  \qc \hat\Theta_0 = 0 \qc \Psi'_{k0} = 0  \qc \hat\Theta_{k0} = 0 .
}
Its numerical solution yields
\eq{29}{
k_0 = 0.709207 \qc \Rm_0 = 10.492375 ,
}
together with the functions $\Psi_0(y)$, $\hat\Theta_0(y)$, $\Psi_{k0}(y)$ and $\hat\Theta_{k0}(y)$.
Furthermore, the $1^{st}$-order differential problem can be written as
\eq{30}{
\Psi_1'' - k_0^2 \Psi_1  - 2 k_0 k_1 \Psi_0 + k_1\, \Rm_0\, \hat\Theta_0 + k_0\, \Rm_1\, \hat\Theta_0 + k_0\, \Rm_0\, \hat\Theta_1 = 0, \nn
\hat\Theta_1'' - k_0^2 \hat\Theta_1 - 2 k_0 k_1 \hat\Theta_0 + \hat\Theta_1' + k_1\, e^{-y}\ \Psi_0 + k_0\, e^{-y}\ \Psi_1 = 0 ,\nn
\Psi_{k1}'' - k_0^2 \Psi_{k1} - 2 k_0 k_1 \Psi_{k0} + k_1\, \Rm_0\, \hat\Theta_{k0} + k_0\, \Rm_1\, \hat\Theta_{k0} + k_0\, \Rm_0\, \hat\Theta_{k1} - 2 k_1\, \Psi_0 \nn
\qquad\qquad- 2 k_0\, \Psi_1 + \Rm_1\, \hat\Theta_0 + \Rm_0\, \hat\Theta_1 = 0, \nn
\hat\Theta_{k1}'' - k_0^2 \hat\Theta_{k1} - 2 k_0 k_1 \hat\Theta_{k0} + \hat\Theta_{k1}' + k_1\, e^{-y}\ \Psi_{k0} + k_0\, e^{-y}\ \Psi_{k1} - 2 k_1\, \hat\Theta_0 - 2 k_0\, \hat\Theta_1 + e^{-y}\ \Psi_1 = 0 ,\nn
y = 0 : \hspace{12mm} \Psi_1 = 0 \qc \hat\Theta_1 = 0 \qc \hat\Theta'_1 = 1 \qc \Psi_{k1} = 0 \qc \hat\Theta_{k1} = 0 \qc \hat\Theta'_{k1} = 0 ,\nn
y \to +\infty : \qquad \Psi_1' = 0  \qc \hat\Theta_1 = 0 \qc \Psi'_{k1} = 0  \qc \hat\Theta_{k1} = 0 .
}
Its numerical solution yields
\eq{31}{
k_1 = 0.072203 \qc \Rm_1 = 2.215147 ,
}
together with the functions $\Psi_1(y)$, $\hat\Theta_1(y)$, $\Psi_{k1}(y)$ and $\hat\Theta_{k1}(y)$. Obviously, one may develop also the $2^{nd}$-order differential problem and so on. The linear approximations of $k_c$ and $\Rm_c$ when $\Bi \ll 1$ yield
\eq{32}{
k_c \approx 0.709207 + 0.072203\, \Bi \qc \Rm_c \approx 10.492375 + 2.215147 \, \Bi .
}
Obviously, one may develop also the $2^{nd}$-order differential problem and, by a similar procedure, higher order approximations, but we will be satisfied with the linear approximation for small Biot numbers.

\begin{figure}[t]
\centering
\includegraphics[height=65mm]{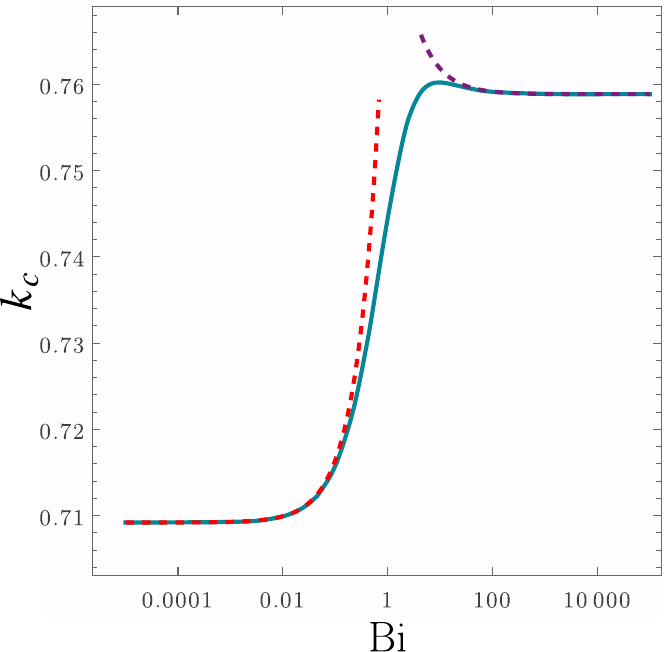}(a)\quad\includegraphics[height=65mm]{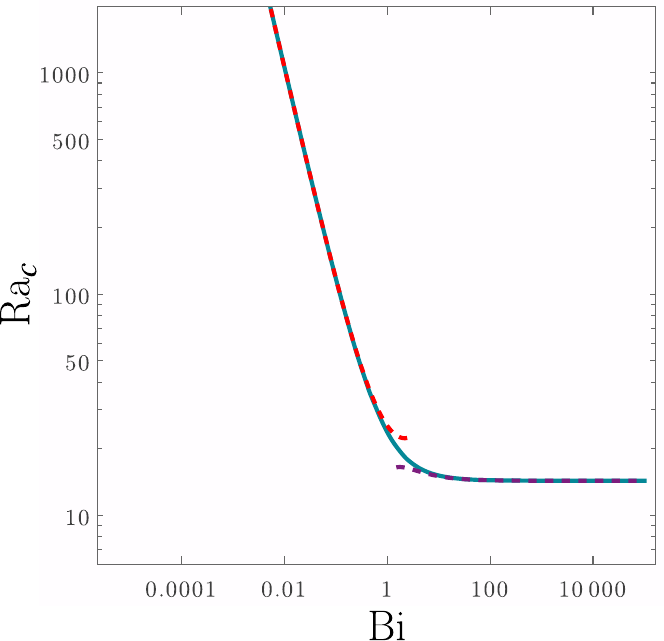}(b)\\[3mm]
\includegraphics[height=65mm]{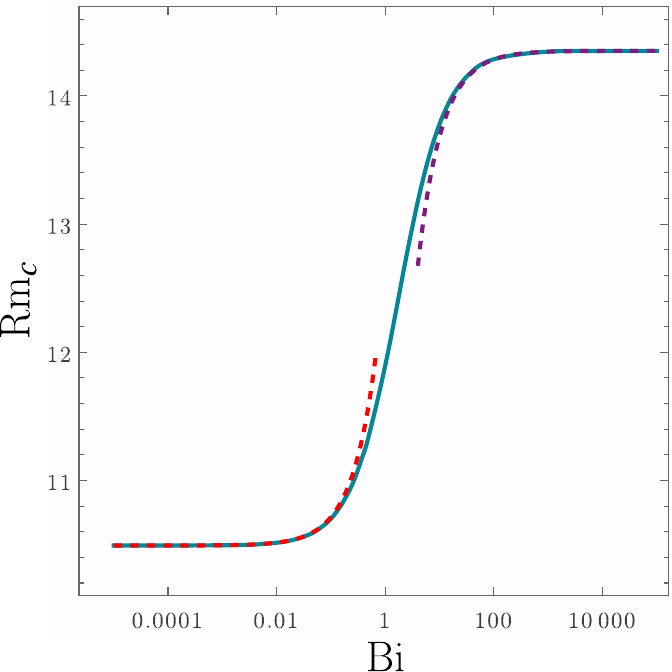}(c)
\caption{\label{fig4}Critical values of $k$, $\Ra$ and $\Rm$ versus $\Bi$. The dashed lines (red and magenta) show the small-$\Bi$ and large-$\Bi$ approximations given by \eqref{32} and \eqref{39}.}
\end{figure}

\begin{table}[t]
\centering
\begin{tabular}{l |r r r }
\hline\hline
\multicolumn{1}{l|}{$\Bi$} & \multicolumn{1}{c}{$k_c$} & \multicolumn{1}{c}{$\Ra_c$} &  \multicolumn{1}{c}{$\Rm_c$} \\
\hline
0     & 0.709207& \multicolumn{1}{c}{$\infty$}    &10.492375 \\
0.001 & 0.709279& 10505.083~~~~~~ 					&10.494589 \\
0.01  & 0.709922& 1061.9542~~~~   					&10.514398 \\
0.02  & 0.710624& 537.34453~~     					&10.536167 \\
0.05  & 0.712653& 222.59995~~     					&10.599998 \\
0.1   & 0.715796& 117.71879~~     					&10.701708 \\
0.2   & 0.721304& 65.335756       					&10.889293 \\
0.5   & 0.733267& 34.054217       					&11.351406 \\
1     & 0.744478& 23.794805       					&11.897402 \\
2     & 0.753847& 18.829919       					&12.553279 \\
5     & 0.759549& 16.026257       					&13.355214 \\
10    & 0.760222& 15.158844       					&13.780767 \\
20    & 0.759900& 14.746259       					&14.044056 \\
50    & 0.759383& 14.507329       					&14.222872 \\
100   & 0.759144& 14.429321       					&14.286456 \\
1000  & 0.758897& 14.359863       					&14.345518 \\
$\infty$  & 0.758867& 14.352191       				&14.352191 \\
\hline
$\infty$,~~ Ref. \cite{homsy1976convective}     & 0.759~~~~~~  & 14.3~~~~~~~~~~       				&\multicolumn{1}{c}{---} \\
$\infty$,~~ Ref. \cite{van2001stability}     & 0.759~~~~~~  & 14.35~~~~~~~~       				&\multicolumn{1}{c}{---} \\
$\infty$,~~ Ref. \cite{rees2009onset}     & 0.7589~~~~  & 14.3552~~~~       				&\multicolumn{1}{c}{---} \\
$\infty$,~~ Ref. \cite{rees2018inclined}  & 0.75887~~~& 14.35219~~       				&\multicolumn{1}{c}{---} \\
$\infty$,~~ Ref. \cite{islam2024onset}    & 0.758867  & 14.352191       				&\multicolumn{1}{c}{---} \\
\hline\hline
\end{tabular}
\caption{\label{tab1}Critical values of $k$, $\Ra$ and $\Rm$ versus $\Bi$.}
\end{table}

\subsubsection{Critical values for large Biot numbers} 
Another approximate solution of \eqref{25} can be obtained when $\Bi \gg 1$.  This task can be accomplished by writing the series expansions in powers of $\tau = 1/\Bi$,
\eq{33}{
\Psi(y) = \Psi_0(y) + \Psi_1(y)\, \tau + \order{\tau^2} \qc \hat\Theta(y) = \hat\Theta_0(y) + \hat\Theta_1(y)\, \tau + \order{\tau^2} , \nn
\Psi_k(y) = \Psi_{k0}(y) + \Psi_{k1}(y)\, \tau + \order{\tau^2} \qc \hat\Theta_k(y) = \hat\Theta_{k0}(y) + \hat\Theta_{k1}(y)\, \tau + \order{\tau^2} , \nn
k = k_0 + k_1\, \tau  + \order{\tau^2} \qc \Rm = \Rm_0 + \Rm_1\, \tau  + \order{\tau^2} .
}
Again, a scale fixing condition can be set which, this time, is chosen as
\eq{34}{
\hat\Theta'(0) = 1 .
}
which implies $\hat\Theta'_0(0) = 1$, $\hat\Theta'_{k0}(0) = 0$, $\hat\Theta'_1(0) = 0$ and $\hat\Theta'_{k1}(0) = 0$, while the boundary conditions in \eqref{25} yield $\hat\Theta_0(0) = 0$, $\hat\Theta_{k0}(0) = 0$, $\hat\Theta_1(0) = 1$ and $\hat\Theta_{k1}(0) = 0$. Thus, substitution of \eqref{33} into \eqref{25} leads to a $0^{th}$-order differential problem, where the four differential equations and the limiting conditions when $y \to +\infty$ are exactly the same given by \eqref{28}, while the boundary conditions at $y=0$ are different,
\eq{35}{
y = 0 : \hspace{12mm} \Psi_0 = 0 \qc \hat\Theta_0 = 0 \qc \hat\Theta'_0 = 1 \qc \Psi_{k0} = 0 \qc \hat\Theta_{k0} = 0 \qc \hat\Theta'_{k0} = 0 .
}
The numerical solution of the $0^{th}$-order differential problem yields
\eq{36}{
k_0 = 0.758867 \qc \Rm_0 = 14.352191 ,
}
together with the functions $\Psi_0(y)$, $\hat\Theta_0(y)$, $\Psi_{k0}(y)$ and $\hat\Theta_{k0}(y)$.

\begin{figure}[t!]
\centering
\includegraphics[height=65mm]{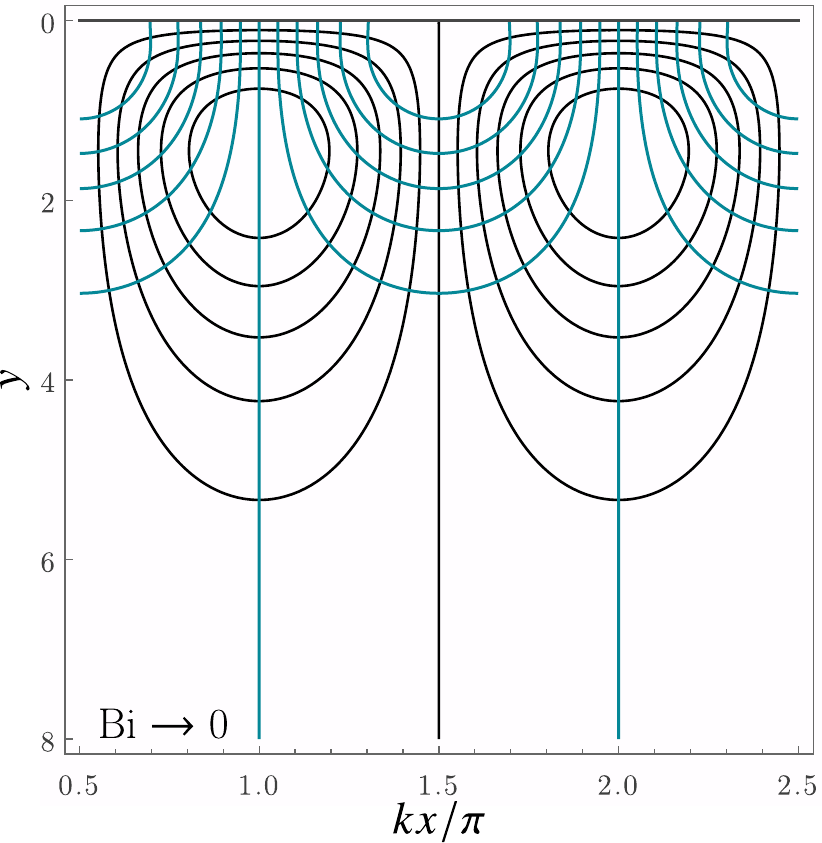}(a)\quad\includegraphics[height=65mm]{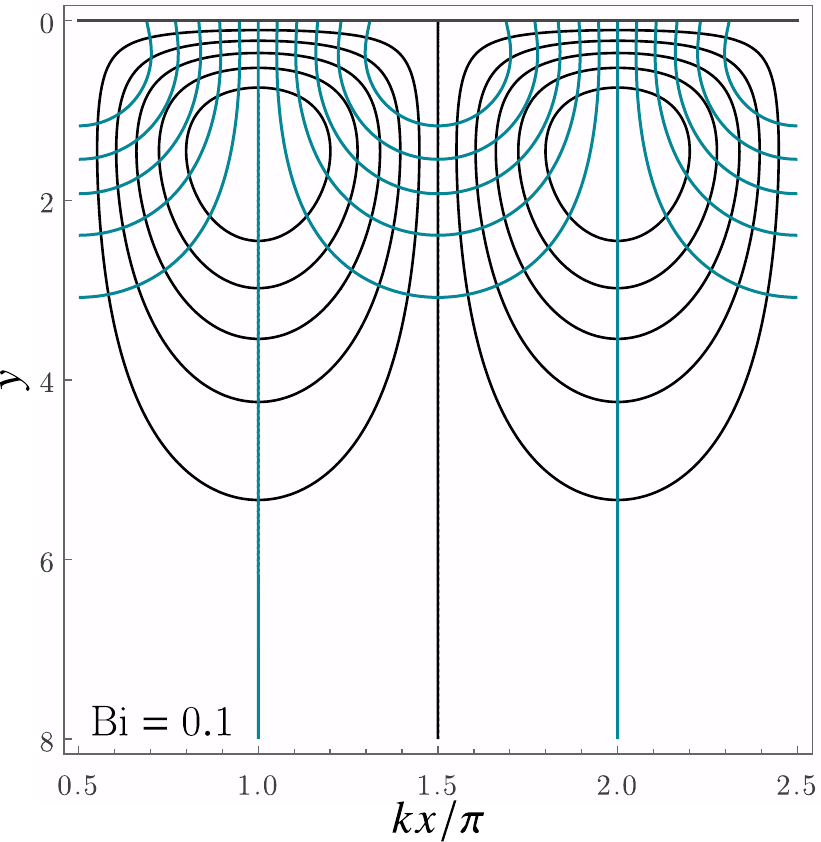}(b)\\[3mm]
\includegraphics[height=65mm]{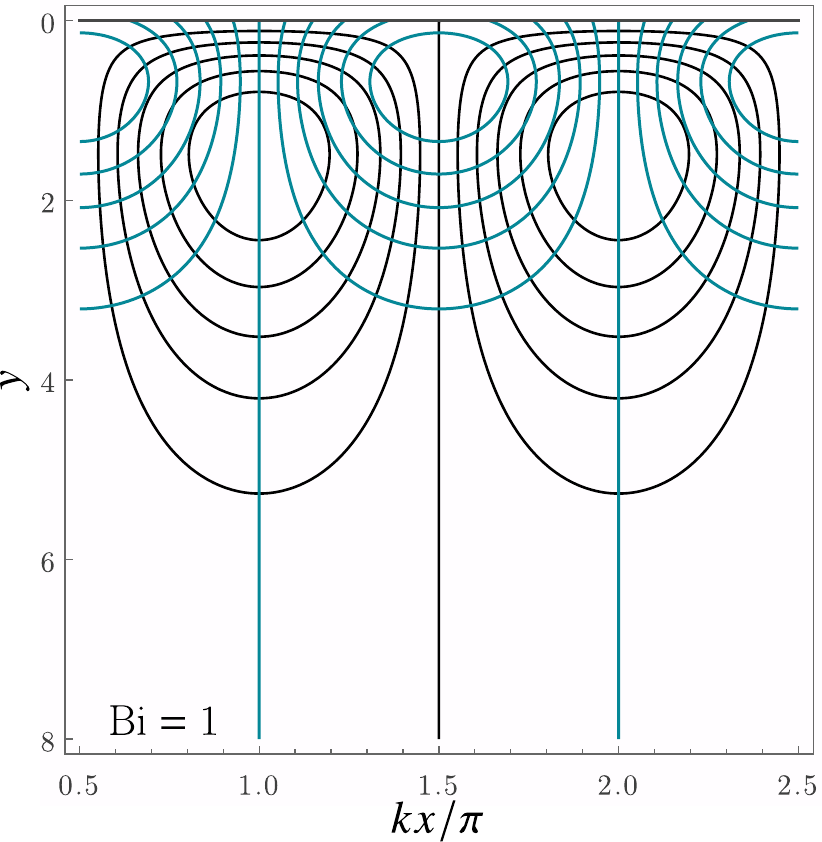}(c)\quad\includegraphics[height=65mm]{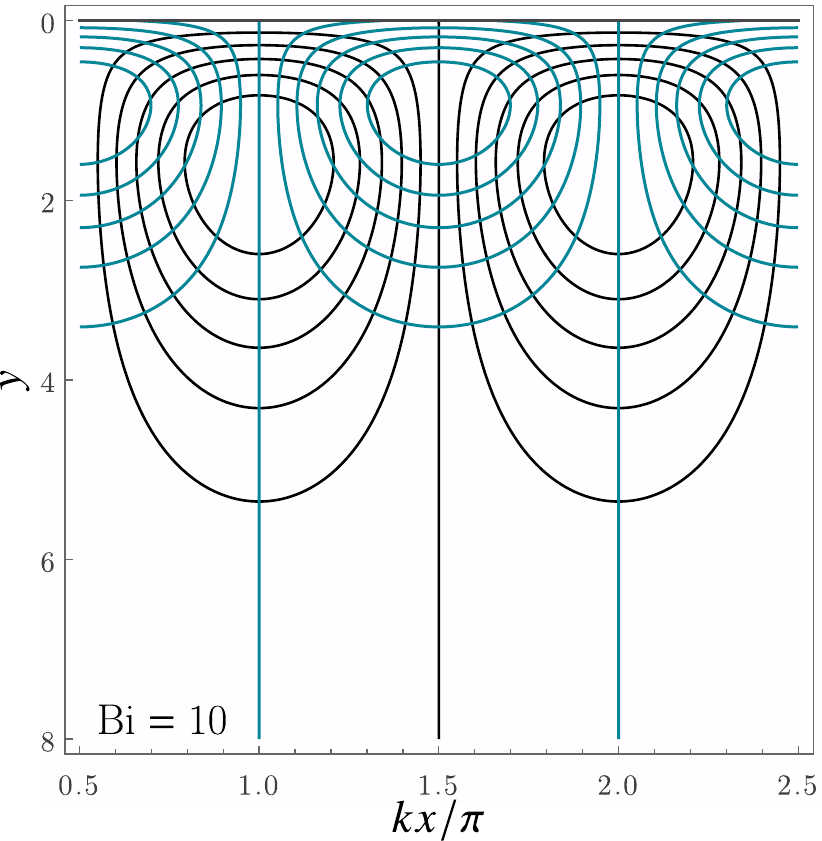}(d)\\[3mm]
\includegraphics[height=65mm]{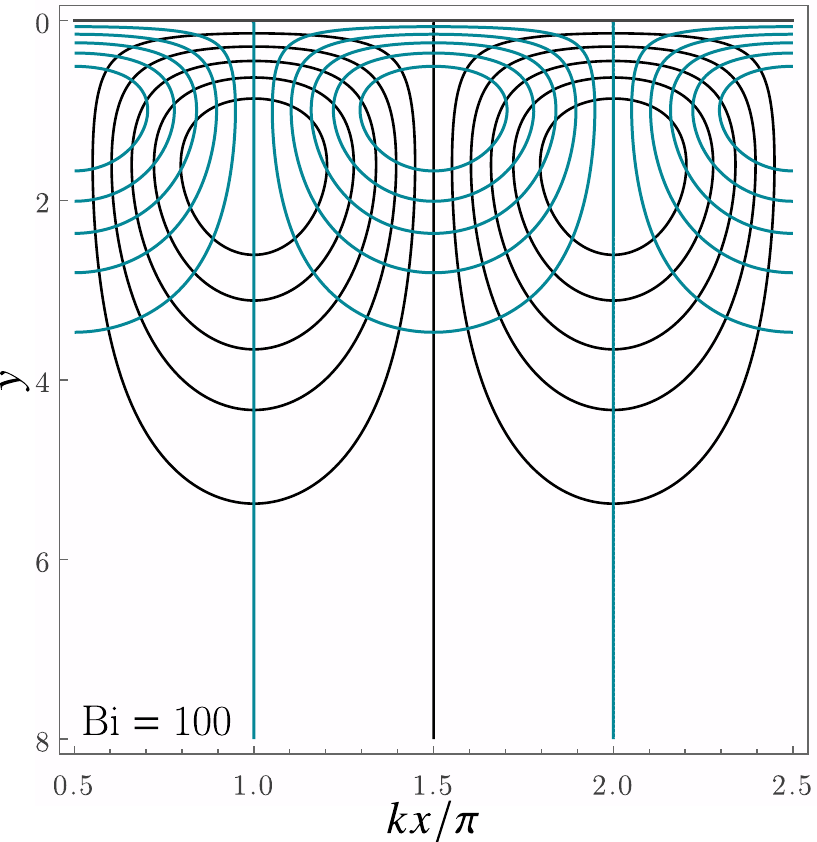}(e)\quad\includegraphics[height=65mm]{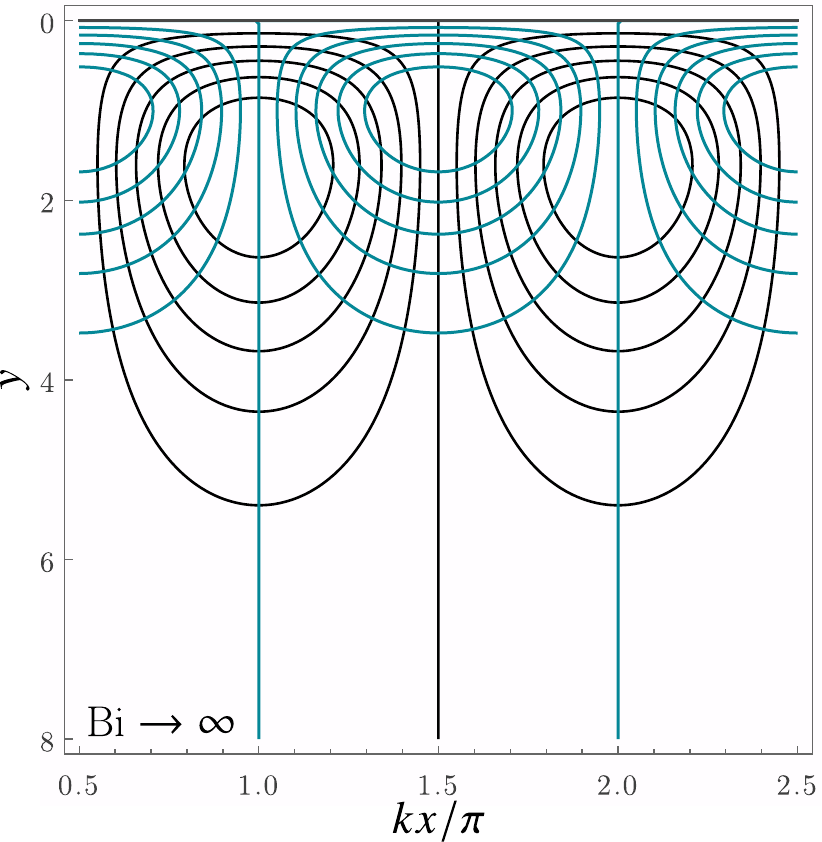}(f)\\[3mm]
\caption{\label{fig5}Perturbation streamlines (black) and isotherms (turquoise) at critical conditions $(k=k_c, \Rm = \Rm_c)$ for increasing Biot numbers.}
\end{figure}

Regarding the $1^{st}$-order differential problem, the four differential equations and the limiting conditions when $y \to +\infty$ are exactly the same as given by \eqref{30}, while the boundary conditions at $y=0$ are different,
\eq{37}{
y = 0 : \hspace{12mm} \Psi_1 = 0 \qc \hat\Theta_1 = 1 \qc \hat\Theta'_1 = 0 \qc \Psi_{k1} = 0 \qc \hat\Theta_{k1} = 0 \qc \hat\Theta'_{k1} = 0 .
}
Its numerical solution yields
\eq{38}{
k_1 = 0.029655 \qc \Rm_1 = -6.684578 ,
}
together with the functions $\Psi_1(y)$, $\hat\Theta_1(y)$, $\Psi_{k1}(y)$ and $\hat\Theta_{k1}(y)$. 
Thus, the approximations of $k_c$ and $\Rm_c$ when $\Bi \gg 1$ can be expressed as
\eq{39}{
k_c \approx 0.758867 + \frac{0.029655}{\Bi} \qc \Rm_c \approx 14.352191 - \frac{6.684578}{\Bi} .
}

\subsubsection{Trend of the critical values versus the Biot number}

The numerical solution of the doubled eigenvalue problem \eqref{25} allows one to plot $k_c$, $\Ra_c$ and $\Rm_c$ versus $\Bi$. Such plots are shown in Fig.~\ref{fig4}, where the dashed lines display the approximated expressions of the critical values for either small or large Biot numbers, given by \eqref{32} and \eqref{39}. We note that the approximated expressions for $\Ra_c$ are obtained from \eqref{32} and \eqref{39} multiplying $\Rm_c$ by $\qty(\Bi + 1)/\Bi$ as a consequence of \eqref{21}. The small-$\Bi$ and large-$\Bi$ approximated formulas turn out to provide a fair approximation of the actual critical data over wide ranges of $\Bi$, with significant discrepancies only when $\Bi$ is around unity, as expected. Figure~\ref{fig4}a shows an apparently larger range of $\Bi$ where the approximated formulas for $k_c$ depart from the numerical data; that is only apparent as it is induced by the tiny vertical range in this plot. Thus, Fig.~\ref{fig4}a highlights the minor change of $k_c$ as $\Bi$ increases and it also displays a very slight maximum of $k_c$ when $\Bi$ is close to $10$. While $\Ra_c$ diverges when $\Bi \to 0$, as shown by Fig.~\ref{fig4}b, the finite limit of $\Rm_c$ when $\Bi \to 0$ is evident in Fig.~\ref{fig4}c. 

Table~\ref{tab1} reports values of $k_c$, $\Ra_c$ and $\Rm_c$ with different Biot numbers. The limiting cases already reported in \eqref{32} and \eqref{39} are also included in this table for completeness. In particular, the critical values for the limit $\Bi \to \infty$ are compared with those reported previously by other authors. An excellent agreement is shown for this limiting case, which is the only one that allows a comparison with the existing literature.

\subsection{Streamlines and isotherms of perturbations}

The perturbations superposed to the basic flow can be visualised, at a given instant of time $t$, by their contour lines in the $(x,y)$ plane. Such a visualisation allows one to appreciate the effects of the gradually changing thermal conditions at $y=0$, as $\Bi$ increases, on the cellular flow and on the temperature distribution. Figure~\ref{fig5} displays the perturbation streamlines and isotherms,
\eq{40}{
\varepsilon\, \psi\qty(x,y,t) = \varepsilon\, \Psi(y)\, e^{\lambda t} \cos(k x) = constant, \nn
\varepsilon\, \frac{\Bi + 1}{\Bi}\, \theta\qty(x,y,t) = \varepsilon\, \hat\Theta(y)\, e^{\lambda t} \sin(k x) = constant,
}
for $k=k_c$ and $\Rm = \Rm_c$ with increasing Biot numbers. The limiting cases $\Bi \to 0$ and $\Bi \to \infty$ are also included. 
The isotherms display an evident change, as $\Bi$ increases, induced by the temperature boundary condition at $y=0$ gradually shifting from a Neumann condition in the limit $\Bi \to 0$ to a Dirichlet condition in the limit $\Bi \to \infty$.

\section{Conclusions}
An extended formulation of the classical Wooding problem has been proposed, where the condition of a perfectly isothermal boundary for the semi-infinite porous medium has been relaxed by imposing a Robin condition for the temperature, parametrised by the Biot number, $\Bi$. A steady-state basic solution has been found where the seepage velocity is uniform and oriented vertically, while a thermal boundary layer decaying exponentially with the distance from the boundary occurs. The linear instability of such a basic solution is analysed by a normal mode analysis. The principle of exchange of stabilities has been proved and a numerical solution of the stability eigenvalue problem has been developed, for different Biot numbers, by utilising the shooting method. Two alternative formulations of the eigenvalue problem have been employed, one based on the classical Rayleigh number for porous media, $\Ra$, and the other based on a suitably rescaled modified Rayleigh number, $\Rm$. The main results of the linear stability analysis can be summarised as follows:
\begin{itemize}
\item The transition to instability identifies opposite roles played by the Biot number. It turned out to be a destabilising parameter if $\Ra$ is employed to parametrise the transition, while it is stabilising if $\Rm$ is employed to parametrise the transition. These opposite roles have been explained physically by recognising that $\Ra$ is a parameter based on a temperature difference, while $\Rm$ is proportional to a heat flux.
\item Both the $\Ra$-parametrisation and the $\Rm$-parametrisation lead to a regular limit $\Bi \to \infty$ for the neutral stability data. Such a limiting case identifies the classical Wooding problem with a Dirichlet boundary condition for the temperature.
\item The limiting case $\Bi \to 0$ leads to a finite neutral stability condition with the $\Rm$-parametrisation, while it predicts linear stability for every finite value of $\Ra$. These seemingly contradictory conclusions are a consequence of the different definitions of $\Ra$ and $\Rm$. If the former parameter is utilised, then $\Bi \to 0$ means an absent basic thermal gradient and, hence, the impossibility of any instability. If the latter parameter is employed, then $\Bi \to 0$ means that the Robin thermal condition is turned into a Neumann condition with a finite heat flux.
\item Low-order perturbative solutions have been obtained for the two regimes $\Bi \ll 1$ and $\Bi \gg 1$. Such solutions and correlations for the critical wave number, and for the critical values of $\Ra$ and $\Rm$ turned out to match satisfactorily the accurate numerical data over wide ranges of $\Bi$.
\end{itemize}
The analysis of the linear instability presented in this paper is a first step to a more complete analysis where nonlinearity will be taken into account. In fact, several authors pointed out the important role of finite-amplitude perturbations in driving the transition to instability \cite{rees2009onset, capone2024throughflow, gianfrani2025eckhaus}. The investigation of a subcritical transition to instability for the case of finite-amplitude perturbations will be the subject of a future paper where the role of an imperfect boundary heat transfer will be studied.

\section*{Acknowledgments}
One of the authors (AB) acknowledges financial support by Alma Mater Studiorum Università di Bologna, grant number RFO-2025.

\end{document}